\documentclass[iop]{emulateapj}

\usepackage{graphicx}
\usepackage{subfigure}
\usepackage{verbatim}
\usepackage{natbib}
\newcommand{\htwo}{H$_2$}

\newcommand{\HII}{H{\small II}}

\newcommand{\acet}{C$_2$H$_2$}
\newcommand{\hcn}{HCN}
\newcommand{\cotwo}{CO$_2$}

\newcommand{\etal}{\textit{et~al.}}

\newcommand{\neII}{[\ion{Ne}{2}]}
\newcommand{\neIII}{[\ion{Ne}{3}]}
\newcommand{\sIII}{[\ion{S}{3}]}
\newcommand{\sIV}{[\ion{S}{4}]}

\newcommand{\neV}{[\ion{Ne}{5}]}

\newcommand{\oIV}{[\ion{O}{4}]}

\newcommand{\um}{$\mu$m}

\shorttitle{MID-IR SPECTROSCOPY OF OH MEGAMASERS: PAPER II}
\shortauthors{WILLETT ET AL.}

\begin{document}

\title{Mid-infrared properties of OH megamaser host galaxies. II. Analysis and modeling of the maser environment}

\author{Kyle W. Willett\altaffilmark{1}, Jeremy Darling\altaffilmark{1}, Henrik W. W. Spoon\altaffilmark{2}, Vassilis Charmandaris\altaffilmark{3,4}, \& Lee Armus\altaffilmark{5}}

\altaffiltext{1}{Center for Astrophysics and Space Astronomy, Department of Astrophysical and Planetary Sciences, UCB 391, University of Colorado, Boulder, CO 80309-0391; willettk@colorado.edu}
\altaffiltext{2}{Astronomy Department, Cornell University, Ithaca, NY 14853}
\altaffiltext{3}{Department of Physics, University of Crete, GR-71003 Heraklion, Greece}
\altaffiltext{4}{IESL/Foundation for Research and Technology-Hellas, GR-71110, Heraklion, Greece; and Chercheur Associ\'e, Observatoire de Paris, F-75014, Paris, France}
\altaffiltext{5}{Spitzer Science Center, California Institute of Technology, Pasadena, CA 91125}


\begin{abstract}
We present a comparison of {\it Spitzer} IRS data for 51 OH megamaser (OHM) hosts and 15 non-masing ULIRGs. 10--25\% of OHMs show evidence for the presence of an AGN, significantly lower than the estimated AGN fraction from previous optical and radio studies. Non-masing ULIRGs have a higher AGN fraction (50--95\%) than OHMs, although some galaxies in both samples show evidence of co-existing starbursts and AGN. Radiative transfer models of the dust environment reveal that non-masing galaxies tend to have clumpy dust geometries commonly associated with AGN, while OHMs have deeper absorption consistent with a smooth, thick dust shell. Statistical analyses show that the major differences between masing and non-masing ULIRGs in the mid-IR relate to the optical depth and dust temperature, which we measure using the 9.7~\um~silicate depth and 30--20~\um~spectral slope from the IRS data. Dust temperatures of $40-80$~K derived from the IRS data are consistent with predictions of OH pumping models and with a minimum $T_{dust}$ required for maser production. The best-fit dust opacities ($\tau_V\sim100-400$), however, are nearly an order of magnitude larger than those predicted for OH inversion, and suggest that modifications to the model may be required. These diagnostics offer the first detailed test of an OHM pumping model based only on the properties of its host galaxy and provide important restrictions on the physical conditions relevant to OHM production. 
\end{abstract}



\section{Introduction}\label{sec-intro}

OH megamasers (OHMs) are 18-cm masers located in the nuclear regions of merging, (ultra)luminous infrared galaxies ([U]LIRGs). Possessing isotropic line luminosities from $10^1-10^4$ $L_\sun$, their hyperfine ratios, extremely broad linewidths, and large physical sizes point to a fundamentally different origin than the Galactic OH masers of the Milky Way \citep{lo05}. A rare phenomenon in the local universe (roughly 100 have been identified out to a redshift of $z=0.265$), OHMs are exceptional probes of their environment due to their ability to be detected at cosmic distances \citep{dar02}. The association of the megamaser emission with merging galaxies means that OHMs trace numerous extreme astrophysical processes, including high-intensity star formation, accretion in the central parts of galaxies, and the eventual formation of massive black holes via binary black hole mergers. 

In order to use OHMs as tracers, however, the relationship between the maser emission and the environment of the host galaxies must be well quantified. Previous studies found no systematic difference between OHM hosts and ULIRGs of similar masses in the radio \citep{lon98,pih05}, optical \citep{baa98,dar06}, or X-ray \citep{vig05} regimes. OHM galaxies do, however, show exceptionally high dense gas fractions and have a distinctly non-linear IR-CO relation \citep{dar07}. Since OHMs are generated deep within the nuclear regions of ULIRGs, however, the maser emission regions are almost always highly obscured, even at near-IR wavelengths. This means that observations capable of probing through the dust are critical both for determining the parameters necessary for production of an OHM and determining its relation with the properties of the host galaxy.  

Mid-IR studies of OHM hosts to date are based primarily on photometry from the IRAS satellite; OHMs tend to occur in galaxies with color excesses at 25 and 60~\um~\citep{hen86}, high IR luminosities \citep{baa89,dar02a}, and steep far-IR spectral indices \citep{che07}.  Spectroscopic studies of the mid-IR emission, however, offer much more powerful diagnostics that can explore the nature of the maser pumping mechanism and the associated OH emission. We used the Infrared Spectrograph (IRS) aboard the \textit{Spitzer Space Telescope} \citep{wer04} to examine the nuclear regions of the merging OHM hosts. Mid-IR observations offer a particularly rich set of diagnostics for ULIRGs, with measurements of AGN activity (high-ionization lines), obscuring dust (absorption features and IR photometry), gas reservoirs (molecular absorption and \htwo~emission), and possible OH reservoirs (hydrocarbons and ices) all visible in the 5--35~\um~region. For some galaxies, the masing gas can also be directly traced via the 34.6~\um~OH transition. 

We recently presented data from a spectroscopic survey of 51 OHM hosts and 15 non-masing galaxies (Willett~\etal~2011; Paper I). In this follow-up analysis paper, we present derived properties of the OHM galaxies and examine statistical differences between the two samples. We also compare physical conditions in the masing regions to those predicted from recent OHM pumping models. Finally, we describe how mid-IR diagnostics may serve as a useful selection technique for future OHM surveys. 


\section{Observations and data reduction}\label{sec-obs}

The {\it Spitzer} data for both the OHMs and the non-masing galaxies come from multiple observing programs; we observed 24 OHMs in a dedicated Cycle 3 program (30407) for IRS observations of OHM hosts. Additional data for both OHMs and confirmed non-masing galaxies were drawn from the Spitzer archive, with approximately half from the IRS GTO sample of ULIRGs. In order to create a uniform data set, we required that all galaxies have full coverage in both the low- and high-resolution modules. Including objects from our dedicated OHM observing program, the samples contain 51 OHM hosts and 15 non-masing galaxies with no OH emission above $L_{OH}=10^{2.3} L_\sun$. 

While OH observations cover the majority of radio galaxies and ULIRGs in the local universe ($z\lesssim0.2$) which are likely candidates for OHM emission, our sample was largely constrained by the availability of data from the {\it Spitzer} archive. This sample is not complete, as there are many objects (both OHMs and non-masing galaxies with firm upper limits) for which mid-IR spectroscopy was not available. 

Since the archival objects did not come from a unified observing program, the version of the \textit{Spitzer} data pipeline and the level of processing vary slightly from object to object - we used the most recent versions available in the archive (v15.3.0 or later). The reduction pipeline is described in detail in Paper~I; briefly, we use the basic calibrated data (BCD) products, subtracting background sky for all LR modules and medianing subsequent exposures to remove transient effects. The images were cleaned using the IDL routine IRSCLEAN\_MASK and the 1-D spectra then extracted using the Spitzer IRS Custom Extractor (SPICE) v.2.0. Almost all galaxies in our sample were unresolved with the IRS and were treated as point sources. 

Low-resolution modules were stitched together to match continuum levels by using a multiplicative scaling, fixing the LL1 module and then stitching the other three modules at the points of overlap. The spectra were calibrated as a single unit to 22 or 25~\um~photometry. Noisy regions (typically 10--30 pixels) corresponding to areas of lowered detector sensitivity at the edges of the SH and LH orders were trimmed from the final 1-D spectra. In isolated cases, exceptional 1-channel features appearing in only a single nod were either manually removed or replaced using the data from the uncorrupted nod position. 


\section{Discussion and analysis}\label{sec-discussion}
	

\subsection{Spectral energy distributions} \label{ssec-cont}	

\begin{figure}
\includegraphics[scale=0.5]{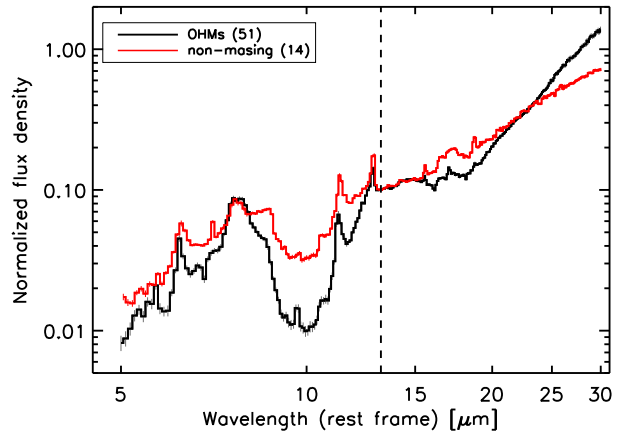}
\caption{Medianed low-resolution spectra for all OHMs (\textit{black}) and non-masing galaxies (\textit{red}). The $1\sigma$ error bars for each medianed pixel are also shown. The dashed line shows where the fluxes are normalized at 15~\um. \label{fig-lravg_both}}
\end{figure}
	
The low-resolution (LR) IRS spectra are powerful indicators of the overall spectral shape of the galaxies, typically dominated by reprocessed emission from dust heated by star formation and/or an active nucleus \citep[e.g.,][]{arm07,hao07}. Figure~\ref{fig-lravg_both} shows the median LR spectra for all galaxies in both the OHM and non-masing samples. A clear difference in the spectral shape between the two samples is apparent; the OHMs show deeper absorption at both 9.7 and 18~\um~and steeper continuum from 15--35~\um. Discrepancies in individual emission and absorption features are also apparent; the PAH emission at 7.7~\um~is broader in the OHM template, with the 8.6~\um~feature largely suppressed (possibly due to extinction from silicate dust). Similarly, the \htwo~S(3)~$\lambda$9.67 line is clearly seen in the median OHM template and suppressed in the non-masing sample. 

	The medianed OHM spectra also reveals a clear absorption feature near 6~\um~associated with water ice \citep{spo02,spo04}; the same feature is not seen in the medianed template of the non-masing galaxies. This is consistent with individual detection rates in the two samples (24/51 OHMs, 3/15 non-masing galaxies). Since water ice is a possible reservoir for the masing OH molecules in their ISM gas-phase, distinct difference between the two populations have implications for OHM emission. If large fractions of the available OH are locked up in solid forms (ice mantles on dust grains, for example), then the reservoir of gas-phase OH could be depleted to a degree that would quench maser emission. This could be due to a harder radiation environment in the non-masing ULIRGs, as sufficiently strong UV radiation can dissociate OH even in the ice phase \citep{and08}. 


\subsection{Narrow-line region gas}

\begin{figure}
\includegraphics[scale=0.5]{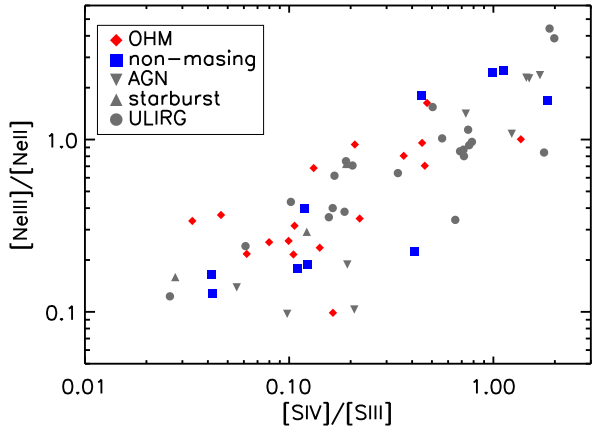}
\caption{Diagnostic of the excitation state in the narrow-line region for OHMs (\textit{red}) and non-masing galaxies ({\it blue}) in which all four of the [Ne~II], [Ne~III], [S~III],  and [S~IV] transitions are detected. Also shown are line ratios for ULIRGs \citep{far07}, active galaxies \citep{stu02,tom08}, and starburst galaxies \citep{ver03}. \label{fig-excitation}}
\end{figure}

	We traced the hardness of the radiation field by comparing the excitation states of the fine-structure neon and sulfur lines, plotting the ratio \neIII/\neII~against the \sIV$\lambda10.5$/\sIII~$\lambda18.71$ (Figure~\ref{fig-excitation}). Detection of all four lines occurred in less than 50\% of the sample (14/51 OHMs, 8/15 non-masing galaxies); non-detection of \sIV~is the limiting factor for almost all galaxies. Since \sIV~lies near the silicate absorption at 9.7~\um, extinction caused by mixing of the dust and ionized gas may suppress observation of this line for dust-rich galaxies. 
	
	The line ratios from our samples are compared to larger populations of ULIRGs \citep{far07}, active galaxies \citep{stu02,tom08}, and starbursts \citep{ver03}. All galaxies show a correlation between higher ionization states for both species, with galaxy types relatively evenly distributed through the total range of line ratios. Fits for each set are consistent with a slope between 0.5 and 1.0; although OHMs have a shallower slope $(0.5\pm0.7$) than the non-masing galaxies ($0.8\pm0.7$), the uncertainties in both fits are too large to distinguish them from each other or the larger population of ULIRGs. These slopes are also consistent with the results of \citet{dal06}, who found that nuclear regions of galaxies in the SINGS sample exhibited a similar trend (although they used \sIII~$\lambda33.48$ instead of \sIII~$\lambda18.71$). Mixing of the line ratios for OHM hosts and non-masing galaxies, along with the lack of a clear locus for either population, suggests that the ionization state of the narrow-line gas is not a factor in triggering an OHM. 

\section{Derived properties}\label{sec-derived}


\subsection{Velocities}\label{ssec-velocities}

\begin{figure}
\includegraphics[scale=0.5]{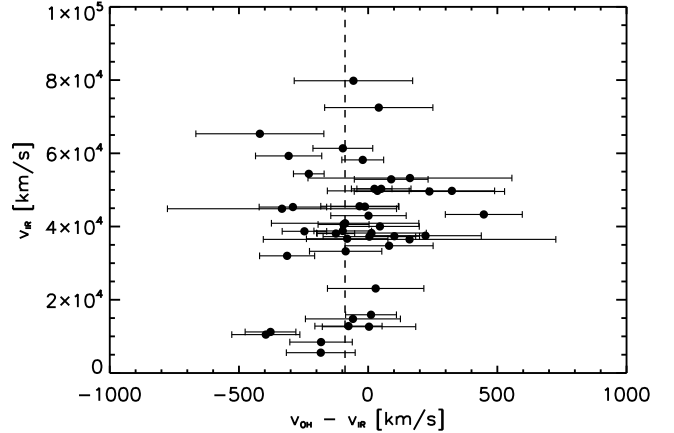}
\caption{OH-IR velocity offset vs. the systemic IR velocity for the OHMs. The mean velocity offset ({\it dashed}) for the sample is $-90\pm19$~km/s, showing a blueshift of the OHM relative to the IR emission lines. Plotted uncertainties are statistical only, and do not account for the unknown velocity ($\lesssim30$~km/s) of the spacecraft at the time of observation. \label{fig-zspitzer_ir_oh}}
\end{figure}

The IRS high-resolution spectra contain multiple narrow lines that can be used for accurate redshift measurements. We computed a systemic IR velocity from the weighted mean of all detected HR line centroids, typically fixed by the strongest transitions (including \htwo, \neII, and \neIII). The scatter of the individual species around the mean velocity is $\sim200$~km/s for the strongest lines, while the velocity resolution of the IRS is $\sim500$~km/s. There is also a possible unknown velocity component from the {\it Spitzer} spacecraft, which may be as high as 30~km/s. We found no statistically significant trend for individual species versus redshift, similar to the trend found in OHM host optical line redshifts by \citet{dar06}.

We compared both the individual and systemic IR velocities to those measured from optical spectroscopy and to the velocity of the OHM itself. \citet{dar06} found a significant asymmetry in the OHM-optical redshift distribution, with the OHM emission somewhat blueshifted with respect to the optical emission. The results for the mean OH-IR velocity offset show a similar blueshift of $\Delta v_{avg} = -90\pm19$~km/s (Figure~\ref{fig-zspitzer_ir_oh}). This is consistent with our measurement of no systematic offset between the IR and optical velocities ($\Delta v_{opt-IR} = -13\pm20$~km/s). While this agrees with the results of \citet{dar07}, the optical/IR agreement is somewhat puzzling given the large amounts of dust in the actively merging galaxies. If the IR lines truly come from the nuclear regions and the optical lines from superficial gas, an offset between the two sets of transitions might be expected - non-detection of this effect (and relatively small scatter) may imply that many of the IR lines are superficial. This is supported by the detection of \htwo~S(3) and \sIV~emission on top of the 9.7~\um~dust absorption feature. We note that for high-ionization lines that must originate near the nucleus (\neV~14 and 24~\um), we do not have enough detections to measure a significant statistical offset. 

The alignment of the mean IR and optical velocities could be partly due to a selection effect, since the lines are primarily identified on the basis of pre-existing optical redshifts (although mis-identification of lines would require offsets of thousands of km/s or greater). \citet{spo09} showed that \neIII~and \neV~emission in ULIRGs can be offset by more than 200~km/s, likely explained by decelerating outflows that are photoionized by AGN. The lack of a systematic blueshift of the OHM in our sample may be an indicator that outflows are not common in the host galaxies, a further indication of a tendency for OHMs not to be associated with AGN. A two-sided Kolmogorov-Smirnov test showed no significant difference in $\Delta v_{opt-IR}$ for the OHM and non-masing populations. 

\citet{dar06} also found a weak correlation between the magnitude of the OHM blueshift and the strength of the OHM (as measured by log~$L_{OH}$ and the linewidth $W_{1667}$). The blueshift of the OHM with respect to the IR emission showed no significant correlation for either parameter for the galaxies in our sample. 


\subsection{Star formation}\label{ssec-sfr}

We also examined the relationship between the OHM and the star formation rate (SFR) in the host galaxies. \cite{ho07} use the fine-structure \neII~and \neIII~lines as diagnostics in galaxies spanning more than five decades of IR luminosity. Neon emission is a useful tracer for the SFR due to its abundance in \HII~regions, ionization energies that make the singly- and doubly-ionized species among the primary coolants for gas heated by massive stars, and relative insensitivity to dust extinction (particularly when compared to common tracers in the optical/UV such as [\ion{O}{3}]~$\lambda5007\AA$ and H$\alpha$).  

\begin{figure}
\includegraphics[scale=0.5]{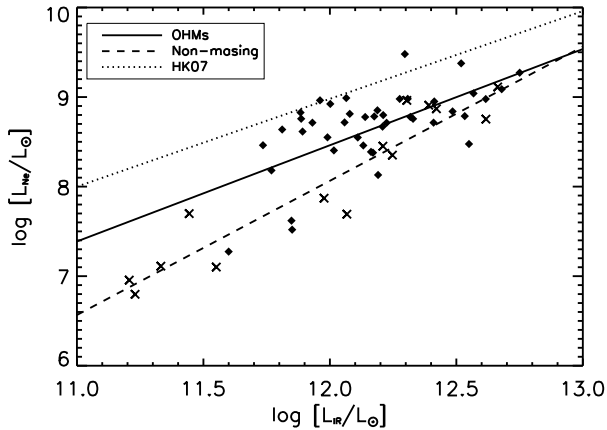}
\caption{Luminosity of [Ne~II]~+~[Ne~III] lines as a function of $L_{IR}$ for the OHM (\textit{diamond}) and non-masing (\textit{cross}) samples. Linear fits to both samples are shown by the solid (OHMs) and dashed (non-masing) lines; the dotted line shows the fit for the much larger sample of \citet{ho07}. The fit to the OHMs is within the scatter of both the other two samples; fits to the HK07 and non-masing galaxies, however, do not agree. \label{fig-sfr_plotII}}
\end{figure}

	Figure~\ref{fig-sfr_plotII} shows the relationship between the far-infrared luminosity \citep[$L_{IR}$; measured using the method of ][]{san96} and the integrated luminosity of the neon lines. We plot the results for both samples in Figure~\ref{fig-sfr_plotII} along with a least-squares linear fit. We also overplot the relations found from the broader sample in \citet{ho07}. Although there is a moderate correlation between $L_{Ne}$ and $L_{IR}$ (as is expected for any comparison involving two luminosities), the scatter is considerable. The combined neon luminosities for the OHMs yield a fit of 

\begin{equation}
\label{eqn-sfr_ohm}
\textrm{log}~[L_{NeII + NeIII}] = (1.0\pm0.5)~\textrm{log}~[L_{IR}] - (4\pm6),
\end{equation}

\noindent with both luminosities measured in $L_\sun$. The fit to the non-masing galaxies is: 

\begin{equation}
\label{eqn-sfr_ohm}
\textrm{log}~[L_{NeII + NeIII}] = (1.5\pm0.4)~\textrm{log}~[L_{IR}] - (10\pm5).
\end{equation}

Both the slope and offset for the OHMs are consistent with the relationship found by \citet{ho07}: log~$[L_{NeII + NeIII}] = (0.98\pm0.069)$~log~[$L_{IR}$] $-$ ($2.78\pm0.70$). The fits for the OHMs and non-masing galaxies are also consistent within the large scatter. The fits to the \citet{ho07} and non-masing galaxies, however, do not overlap (within their respective 1$\sigma$ scatter), indicating a possible marginal difference in SFR.   

	The larger uncertainties in the OHM and non-masing galaxies' slopes are attributed to their narrow range in $L_{IR}$. Both samples have a lower limit on $L_{IR}$ that lies at the high end of the \citet{ho07} data. The upper end of the $L_{IR}$ range reflects the low space density of HyLIRGs with $L_{IR}>10^{12.5}~L_\sun$. The net effect yields only $\sim1.5$~dex of $L_{IR}$ over which a relation can be fit; since the galaxies in \citet{ho07} is over more than five decades of $L_{IR}$, their correlation is much tighter. \citet{far07} suggest the offset between the slopes is a result of higher extinction in the nuclear regions of ULIRGs relative to lower-luminosity starbursts that fix the height of the \citet{ho07} relation. The fact that both samples are neon-underluminous compared to the larger data set agrees with the high extinction ($\tau_V\sim300$) found by fitting dust models (\S\ref{ssec-dustgeometry}). Since the non-masing galaxies have even lower neon fluxes, their total $L_{IR}$ likely has a lower overall contribution from star formation. 

\begin{deluxetable*}{llrrrrcrccrc}
\tabletypesize{\scriptsize}
\tablecaption{Derived mid-IR and radio properties for OHMs and non-masing galaxies\label{tbl-derived}}
\tablewidth{0pt}
\tablehead{
\colhead{} & 
\colhead{} & 
\multicolumn{2}{c}{\underline{Star formation rates}} &
\colhead{} &
\multicolumn{4}{c}{\underline{DUSTY best-fit parameters}} &
\colhead{} &
\multicolumn{2}{c}{\underline{LE08 predictions}}
\\
\colhead{} & 
\colhead{Object} & 
\colhead{$SFR_{Ne}$} &
\colhead{$SFR_{FIR}$} &
\colhead{$T_{dust}^{gb}$} & 
\colhead{$Y$} & 
\colhead{$q$} & 
\colhead{$\tau_V$} & 
\colhead{$T_{dust}$} &
\colhead{$\tau_{1667}^{app}$} &
\colhead{$\tau_{1667}^{IRS}$} &
\colhead{$\tau_{1667}^{DUSTY}$}
\\
\colhead{} & 
\colhead{} & 
\colhead{[M$_\sun$/yr]} &
\colhead{[M$_\sun$/yr]} &
\colhead{[K]} &
\colhead{} &
\colhead{} &
\colhead{} &
\colhead{[K]} &
\colhead{} &
\colhead{} &
\colhead{}
}
\startdata
OHMs       & IRAS 01355$-$1814 & 163 &  502 & 60  &  250  & 0.0  & 410  &  64   &  --	  & $-1.0$	& **	\\
           & IRAS 01418+1651   &   3 &   69 & 54  &  250  & 0.0  & 410  &  64   &  $-$1.9 & $-1.0$ 	& **	\\
           & IRAS 01562+2528   &  64 &  250 & 57  &  950  & 0.0  &  55  &  43   &  $-$0.7 & $ 0.0$ 	& 0.0	\\
           & IRAS 02524+2046   &  -- &  204 & 62  &  350  & 0.0  & 410  &  55   &  $-$2.7 & $-1.0$ 	& **	\\
           & IRAS 03521+0028   & 111 &  595 & 62  &  250  & 0.0  & 410  &  64   &  $-$0.3 & $-1.5$ 	& **	\\
           & IRAS 04121+0223   &  -- &   80 & 63  &  300  & 0.0  & 300  &  61   &  $-$0.6 & $-1.0$ 	& $-1.0$	\\
           & IRAS 04454$-$4838 &   7 &  122 & 64  &  300  & 0.0  & 410  &  59   &    --   & $-1.5$      &  **   \\
           & IRAS 06487+2208   & 551 &  384 & 76  &  200  & 0.0  & 190  &  74   &  $-$0.5 & $-2.0$ 	& $-1.5$	\\
           & IRAS 07163+0817   &  52 &  105 & 61  &  250  & 0.0  & 310  &  65   &  $-$0.8 & $-1.0$ 	& **	\\
           & IRAS 07572+0533   &  -- &  350 & 81  & 1000  & 0.0  &  46  &  43   &  $-$0.4 & $-2.0$ 	& 0.0	\\
           & IRAS 08201+2801   & 114 &  315 & 71  &  200  & 0.0  & 410  &  70   &  $-$0.6 & $-2.0$ 	& **	\\
           & IRAS 08449+2332   & 153 &  194 & 70  &  200  & 0.0  & 340  &  71   &  $-$0.3 & $-1.5$ 	& **	\\
           & IRAS 08474+1813   &  34 &  251 & 61  &  250  & 0.0  & 410  &  64   &  $-$0.4 & $-1.5$ 	& **	\\
           & IRAS 09039+0503   &  95 &  221 & 58  &  300  & 0.0  & 290  &  61   &  $-$0.6 & $ 0.0$ 	& $-1.0$	\\
           & IRAS 09539+0857   &  -- &  190 & 49  &  250  & 0.0  & 380  &  64   &  $-$0.9 & $-2.5$ 	& **	\\
           & IRAS 10035+2740   &  85 &  313 & 58  &  300  & 0.0  & 410  &  59   &  $-$0.3 & $-1.0$ 	& **	\\
           & IRAS 10039$-$3338 &  27 &   88 & 71  & 1000  & 1.0  & 120  &  37   &  $-$2.6 & $-2.5$ 	& 0.0	\\
           & IRAS 10173+0828   &   6 &  108 & 51  &  300  & 0.0  & 410  &  59   &  $-$2.4 & $-1.0$ 	& **	\\
           & IRAS 10339+1548   & 174 &  394 & 63  &  250  & 0.0  & 380  &  64   &  $-$0.8 & $-1.0$ 	& **	\\
           & IRAS 10378+1109   & 109 &  348 & 69  &  200  & 0.0  & 410  &  70   &  $-$0.9 & $-2.0$ 	& **	\\
           & IRAS 10485$-$1447 &  52 &  263 & 64  &  250  & 0.0  & 410  &  64   &  --     & $-1.5$ 	& **	\\
           & IRAS 11028+3130   &   0 &  420 & 55  &  250  & 0.0  & 410  &  64   &  $-$0.6 & $-1.0$ 	& **	\\
           & IRAS 11180+1623   &  94 &  325 & 62  &  250  & 0.0  & 410  &  64   &  $-$0.4 & $-1.5$ 	& **	\\
           & IRAS 11524+1058   &  -- &  268 & 58  &  350  & 0.0  & 250  &  58   &  $-$0.5 & $-1.0$ 	& $-1.0$	\\
           & IRAS 12018+1941   & 126 &  454 & 74  &  150  & 0.0  & 200  &  83   &  $-$0.4 & $-3.0$ 	& $-1.5$	\\
           & IRAS 12032+1707   & 434 &  641 & 94  &  200  & 0.0  & 410  &  70   &  $-$0.4 & $-2.0$ 	& **	\\
           & IRAS 12112+0305   & 104 &  372 & 55  &  300  & 0.0  & 410  &  59   &  $-$1.1 & $-0.5$ 	& **	\\
           & IRAS 12540+5708   &  54 &  419 & 77  &  950  & 0.5  &  50  &  43   &  $-$0.1 & $-2.5$ 	& 0.0	\\
           & IRAS 13218+0552   &  -- &  415 & 95  &  950  & 1.5  &  56  &  41   &  $-$0.6 & $-3.0$ 	& 0.0	\\
           & IRAS 13428+5608   & 110 &  244 & 65  &  250  & 0.0  & 200  &  67   &  $-$0.4 & $-1.5$ 	& $-1.5$	\\
           & IRAS 13451+1232   & 174 &  253 & 68  & 1000  & 0.0  &  34  &  44   &  $-0.0003$ & $-2.0$ 	& 0.0	\\
           & IRAS 14059+2000   & 104 &  149 & 63  &  950  & 0.0  &  61  &  43   &  $-$1.1 & $-0.5$ 	& 0.0	\\
           & IRAS 14070+0525   & 341 & 1092 & 95  &  250  & 0.0  & 410  &  64   &  $-$1.1 & $-1.5$ 	& **	\\
           & IRAS 14553+1245   & 122 &  128 & 73  &  250  & 0.0  & 190  &  68   &  $-$0.6 & $-2.0$ 	& $-1.5$	\\
           & IRAS 15327+2340   &  24 &  266 & 60  &  250  & 0.0  & 410  &  64   &  $-$0.6 & $-1.5$ 	& **	\\
           & IRAS 16090$-$0139 & 200 &  607 & 62  &  300  & 0.0  & 190  &  63   &  $-$0.4 & $-1.5$ 	& $-1.5$	\\
           & IRAS 16255+2801   &  64 &  151 & 70  &  250  & 0.0  & 210  &  67   &  $-$0.9 & $-2.0$ 	& $-1.5$	\\
           & IRAS 16300+1558   & 223 &  927 & 54  &  300  & 0.0  & 410  &  59   &  $-$0.3 & $-1.5$ 	& **	\\
           & IRAS 17207$-$0014 &  94 &  456 & 59  &  300  & 0.0  & 410  &  59   &  $-$1.0 & $-1.5$ 	& **	\\
           & IRAS 18368+3549   & 129 &  299 & 56  &  300  & 0.0  & 260  &  61   &  $-$0.2 & $ 0.0$ 	& $-1.0$	\\
           & IRAS 18588+3517   &  94 &  144 & 71  &  250  & 0.0  & 280  &  66   &  $-$0.8 & $-2.0$ 	& $-1.0$	\\
           & IRAS 20100$-$4156 & 173 &  732 & 62  &  250  & 0.0  & 410  &  64   &    --   & $-1.0$      & **    \\
           & IRAS 20286+1846   &  46 &  201 & 37  &  250  & 0.0  & 230  &  67   &  $-$1.4 & $ 0.0$ 	& $-1.5$	\\
           & IRAS 21077+3358   & 167 &  177 & 76  &  200  & 0.0  & 410  &  70   &  $-$0.4 & $-2.5$ 	& **	\\
           & IRAS 21272+2514   &  75 &  151 & 77  &  200  & 0.0  & 410  &  70   &  $-$1.5 & $-2.0$ 	& **	\\
           & IRAS 22055+3024   & 109 &  267 & 71  &  200  & 0.0  & 240  &  73   &  $-$0.7 & $-2.0$ 	& $-1.5$	\\
           & IRAS 22116+0437   & 178 &  225 & 76  &  200  & 0.0  & 380  &  71   &  $-$0.2 & $-2.5$ 	& **	\\
           & IRAS 22491$-$1808 &  44 &  246 & 58  &  250  & 0.0  & 410  &  64   &  $-$1.1 & $-1.5$ 	& **	\\
           & IRAS 23028+0725   &  79 &  125 & 91  &  --   & --   & --   &  --   &  --     & --          & --    \\
           & IRAS 23233+0946   & 118 &  232 & 67  &  200  & 0.0  & 380  &  71   &  $-$0.3 & $ 0.0$	& **	\\
           & IRAS 23365+3604   &  43 &  245 & 70  &  200  & 0.0  & 410  &  70   &  --     & $-2.0$	& **	\\
\hline                                                                       
Non-masing & IRAS 00163$-$1039 &  69 &   40 & 52  & 1000  & 0.0  &  50  &  43   &  --	& $ 0.0$	& 0.0	\\
           & IRAS 01572+0009   & 559 &  513 & 81  &  850  & 0.0  &  23  &  48   &  --	& $-0.5$ 	& 0.0	\\
           & IRAS 05083+7936   & 181 &  148 & 43  & 1000  & 0.0  &  34  &  44   &  --	& $ 0.0$	& 0.0	\\
           & IRAS 06538+4628   &  26 &   30 & 51  &   75  & 0.0  & 380  & 109   &  --	& $ 0.0$	& **	\\
           & IRAS 08559+1053   & 288 &  261 & 74  & 1000  & 0.5  &  46  &  42   &  --	& $-1.5$	& 0.0	\\
           & IRAS 09437+0317   &   4 &   24 & 38  & 1000  & 0.0  &  46  &  43   &  --	& $ 0.0$	& 0.0	\\
           & IRAS 10565+2448   & 139 &  188 & 55  &  950  & 0.0  &  67  &  43   &  --	& $-0.5$ 	& 0.0	\\
           & IRAS 11119+3257   & 196 &  519 & 78  &  150  & 1.0  &  51  &  82   &  --	& $-2.0$	& $-2.5$	\\
           & IRAS 13349+2438   &  69 &   42 & 244 &  950  & 2.0  &  90  &  40   &  --	&   --   	& 0.0	\\
           & IRAS 15001+1433   & 318 &  454 & 67  & 1000  & 0.0  &  55  &  43   &  --	& $-1.5$	& 0.0	\\
           & IRAS 15206+3342   & 591 &  237 & 78  & 1000  & 0.0  &  34  &  44   &  --	& $-1.5$	& 0.0	\\
           & IRAS 20460+1925   &  -- &  186 & 112 & --    & --   &  --  &  --   &  --	&   --   	& --	\\
           & IRAS 23007+0836   &  50 &   45 & 79  &  650  & 0.0  &  28  &  53   &  --	& $-1.5$	& 0.0	\\
           & IRAS 23394$-$0353 &  28 &   22 & 44  &  700  & 0.0  &  56  &  49   &  --	& $ 0.0$	& 0.0	\\
           & IRAS 23498+2423   & 662 &  477 & 93  &  500  & 1.0  &  45  &  53   &  --	& $-3.0$	& 0.0	\\
\enddata
\tablecomments{$T_{dust}^{gb}$ is the dust temperature fit to the photometric greybody; the second $T_{dust}$ is the temperature at the outer envelope fit by the DUSTY models. $\tau_{1667}^{app}$ is the apparent maser optical depth calculated from radio fluxes in the literature (Equation~\ref{eqn-tauohapp}), while $\tau_{1667}^{IRS}$ and $\tau_{1667}^{DUSTY}$ are predicted values from the LE08 model based on the data in \S\ref{ssec-le08v1} and \S\ref{ssec-le08v2}, respectively. ** indicates that the predicted $\tau_{1667}$ fell outside the available contours for the LE08 model.}
\end{deluxetable*}

	We computed star formation rates for our samples using two diagnostics: $L_{Ne}$ from \citet{ho07} and the starburst far-IR luminosity calibration from \citet{ken98a} (Table~\ref{tbl-derived}). For the neon relation, we assume an ionization fraction of $f_{ion}=0.6$ and neon ionization fractions of $f_{Ne^+} = 0.75$ and $f_{Ne^{++}} = 0.1$. The OHM mean star formation rate is $\langle SFR_{Ne} \rangle=120\pm16~M_\sun$/yr, compared to $\langle SFR_{Ne} \rangle=65\pm21~M_\sun$/yr for non-masing galaxies. Using the $L_{FIR}$ calibration with IRAS photometry yields a higher average SFR and larger scatter for both samples, with $\langle SFR_{FIR} \rangle=300\pm30~M_\sun$/yr for OHMs and $\langle SFR_{Ne} \rangle=210\pm50~M_\sun$/yr for non-masing galaxies. The closer agreement between the two populations for this calibrator reflects the $L_{IR}$ criterion used in selecting the non-masing sample.  

	Both the large scatter and the difference in SFR between the diagnostics illustrate the difficulties in characterizing a local phenomenon over a large volume. This emphasizes the fact that all OHMs are mergers with multiple components, with sites of star formation likely separated by tens of kpc. The linear correlation between the two SFRs is also relatively low, with a Spearman's rho of $\rho=0.54$. This could indicate a component for heating the dust that does not come from star formation, such as an AGN.

	

\subsection{AGN vs. starburst}\label{ssec-agn}

	A key issue surrounding ULIRGs is their central source of power - does it come from AGN or starbursts? \cite{baa98} use optical classifications of OHM hosts to claim that 45\% of the host galaxies show signs of a pure AGN (Seyfert and LINER spectra), with an additional 22.5\% displaying composite spectra with characteristics from both AGN and starburst activity. \cite{dar06} compare a sample of OHM host galaxies vs. non-masing ULIRGs and find that 42\% are LINERs, 25\% Seyfert 2 galaxies, and 33\% starbursts; classifications are similar for both samples. They also find few significant correlations between the OHM emission and the optical properties of their host galaxies. Classification using the radio and FIR properties of the nuclei, however, show only 34\% of the sample with AGN characteristics \citep{baa06}; these include multiple objects optically classified as LINERs or composite objects that show no AGN activity in the radio. It is suggested that the differences in classification lie in the large amounts of extinction at optical wavelengths due to dust obscuring the nucleus. 
	
	In the mid-IR, high-ionization fine structure emission lines are the simplest and most unambiguous tracers of AGN activity. \neV~has an ionization energy of 97.1~eV, a level typically too high to be reached by young O and B stars. \oIV~has a smaller ionization energy of 55~eV, which is often seen in AGN and in several optically-identified starburst galaxies. In contrast, the \neV~line is \emph{only} seen in integrated galactic spectra that harbor AGN, although it is not ubiquitous  - \cite{dud07} detect \neV~in 19/41 Seyfert and LINER galaxies, for example. It is possible in these cases that lines are present, but that differential extinction in the mid-IR obscures their emission. Since AGN occupy a much smaller volume ($<~1$~pc) than a typical starburst and have harder radiation fields, the high-ionization regions where the neon and oxygen are emitted are likely to be more deeply obscured than their low-ionization counterparts. 
	
\neV~$\lambda14.3$ is detected in 4/51 OHMs and 8/15 non-masing galaxies (Paper~I). ULIRGs in the larger sample of \citet{far07} show \neV~$\lambda14.3$ in 22/53 galaxies, three of which are OHMs that overlap with our sample. \oIV~is seen in 21/22 galaxies in their sample that display \neV~and only in two that do not, demonstrating a close but not perfect association. The difference in detection rates suggests that the presence of an OHM selects against AGN with high-ionization emission; this may relate to the timescale of the galactic merger. If OHMs are associated with a particular phase since the onset of the host galaxies' merger (and possibly a delay before the activation of the AGN), then this would explain why the OHM sample has so few \neV~detections compared to non-masing galaxies and ULIRGs in general. 
	
	\begin{figure}
	\includegraphics[scale=0.5]{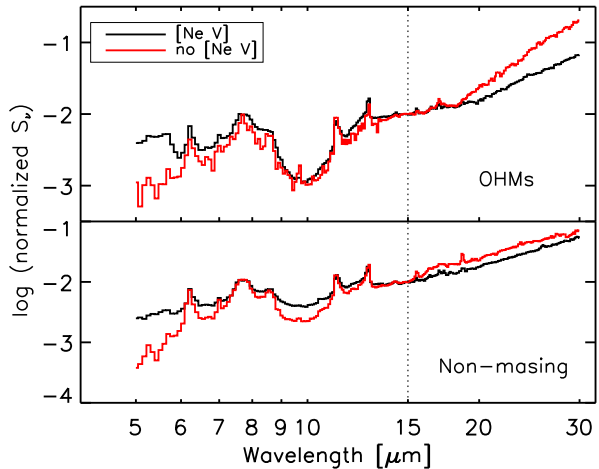}
	\caption{IRS low-resolution spectra, showing the difference in the median-stacked spectra for galaxies with [Ne~V] emission ({\it black}) and with [Ne~V] upper limits ({\it red}). \neV~was detected in 4/51 OHMs ({\it top}) and 8/15 non-masing galaxies ({\it bottom}). Spectra are normalized in flux at $\lambda=15$~\um~({\it dotted line}).\label{fig-nev}}
	\end{figure}

	Given the strong differences in the \neV~detection rate between the samples, we examined whether galaxies emitting \neV~might reveal other parameters relevant to OHM formation. In the average IRS LR spectra for OHMs (Fig.~\ref{fig-nev}), galaxies with \neV~emission show a shallower 30--20~\um~slope than galaxies without high-ionization lines ($\alpha_{30-20}=3.7$ vs. 5.4). The silicate depths and PAH luminosities, however, are broadly consistent for both samples. For non-masing galaxies, Figure~\ref{fig-nev} shows that the average 9.7~\um~silicate depth is shallower for galaxies that show \neV~($S_{9.7}=0.5$ vs 0.7), but that the $\alpha_{30-20}$ slopes are similar. The EW of the 6.2~\um~PAH feature is also smaller in galaxies with \neV; the high-ionization lines are consistent with the presence of an AGN that dissociates large molecules. For the OHMs, the presence of \neV~shows no effect on either $L_{OH}$ or the peak flux at 1667~MHz.

	Other mid-IR diagnostics can also be used to characterize the contribution of AGN and/or starburst features. \cite{spo07} plot the PAH 6.2~\um~EW against the silicate 9.7~\um~strength in a ``fork'' diagram. IRS data show two distinct branches of galaxies for this diagnostic, with one representing a largely AGN-dominated population (weak PAH emission and little to no silicate absorption) and another containing ULIRGs/HyLIRGs, obscured AGN, and starburst galaxies (stronger PAH emission coupled with deeper silicate strengths). 
	
\begin{figure}
\includegraphics[scale=0.5]{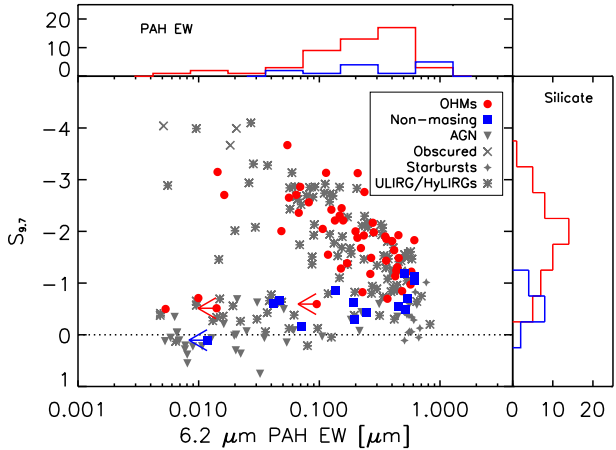}
\caption{``Fork'' diagram displaying the 6.2~\um~PAH equivalent width vs. the silicate strength at 9.7~\um. OHM galaxies from both our program and the archive are shown in red; non-masing galaxies are shown in blue. \textit{Top}: binned distribution of PAH EW for both samples; \textit{Right}: binned distribution of $S_{9.7}$. Additional \textit{Spitzer} data are from \citet{spo07}. \label{fig-spoon_forkII}}
\end{figure}

	We reproduce the fork diagram from \cite{spo07} with our IRS data overlaid on the broader sample of ULIRGs, starburst galaxies, and AGN in Figure~\ref{fig-spoon_forkII}. OHMs lie almost exclusively along the top branch and share significant overlap with optically identified starburst galaxies, which typically have strong PAH emission but weak to moderate silicate absorption. The locus of the OHMs on the fork diagram agrees with the \neV~and \oIV~data; only four OHMs lie on the horizontal, AGN-dominated branch. The non-masing galaxies are principally found along the horizontal branch, with a wide range of PAH EW but lower $S_{9.7}$ than the OHMs. A small region of overlap does exist between the two samples near the ``knee'' (high PAH EW and weaker silicate absorption). The absence of non-masing ULIRGs on the upper branch is one of the first clear spectral diagnostics of OHMs, based only on the properties of the host galaxy.

	Using the mid-infrared diagnostics, we estimate the AGN contribution by assuming that \neV~clearly indicates an AGN and that \oIV~detection or placement on the horizontal branch in the fork diagram indicates a possible AGN. Based on the IRS data, the AGN fraction of OHMs is between 10 and 25\%, compared to a much higher fraction of 50--95\% for non-masing ULIRGs. The AGN contributions for the combined samples are consistent with that estimated from radio/FIR diagnostics \citep{baa06}. The co-existence of AGN and starbursts in some nuclei is also supported by the presence of galaxies showing both high-ionization emission and large PAH equivalent widths. The mid-IR AGN fraction for OHMs, however, is significantly lower than that estimated from optical diagnostics  \citep[45--70\%;][]{baa98}, which can be significantly affected by dust obscuration around the nuclear regions.  

\subsection{Dust temperatures}\label{ssec-dusttemp}

	We computed dust temperatures for both samples using two methods. The first is a single fit to the integrated IR emission by assuming a single-temperature modified blackbody as a template (the second is taken from the model fits in \S\ref{ssec-dustgeometry}). We adapt the broad SED of \cite{yun02}, where the emission follows a thermal blackbody above a critical frequency $\nu_c$ where the dust clouds become optically thick and a greybody spectrum below $\nu_c$. For an object subtending an angular diameter $\theta$~[arcsec], the expected flux density at frequency $\nu$~[GHz] is
	
\begin{equation}
\label{eqn-dusttemplate}
S_d [\nu] = 2.8\times10^{-8} \frac{\nu^3 \theta^2}{e^{0.048\nu/T_d} - 1} \left(1-e^{{(\nu/\nu_c)}^\beta}\right) \textrm{ Jy}.
\end{equation}

	In our analysis, $\theta$ is a free parameter (accounting for the dependence of flux density on distance) and assumed $\nu_c=2000$~GHz (150~\um) and an emissivity index $\beta=1.35$. The fits are relatively insensitive to the choice of $\beta$ since for most galaxies we lack photometric data points below the critical frequency. We used fluxes from IRAS and IRS peakups to fit the curve with photometry from 12--100~\um, allowing both the physical temperature and the peak intensity (a function of both distance and extinction) to vary. The majority of the galaxies only have IRAS detections at 60 and/or 100~\um, in addition to the IRS peakups. 
	
	The mean temperature for the OHMs is $T_{dust}=66\pm12$~K, while the non-masing galaxies have $T_{dust}=80\pm50$~K. These uncertainties are the statistical 1$\sigma$ envelopes for the sample and do not address the physical relevance of fitting the galaxies with a single temperature fit. The hottest temperature measured is in the non-masing galaxy IRAS~13349+2438 ($T_{dust}=243$~K), which is more than twice as hot as the next-highest galaxy. The greybody temperatures for all galaxies are listed as $T_{dust}^{gb}$ in Table~\ref{tbl-derived}. 


\subsection{Modeling the dust environment}\label{ssec-dustgeometry}

	\begin{figure}
	\includegraphics[scale=0.5]{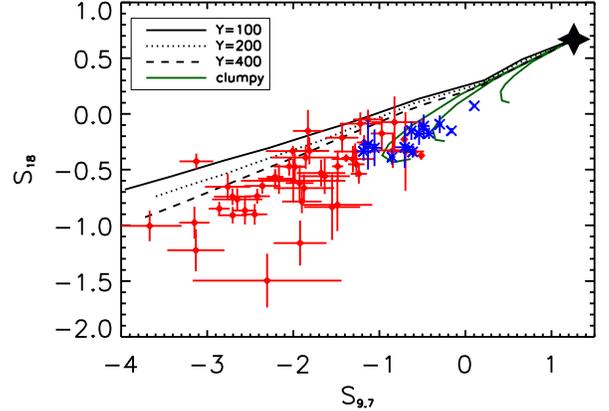}
	\caption{Feature-feature diagram plotting the relative strengths of the 9.7 and 18~\um~silicate features. The lines represent models of different dust geometries for cool, oxygen-rich silicates \citep{sir08}. Dotted, dashed, and solid lines represent the thickness of the dust shell in the smooth models ($Y=R_{outer}/R_{inner}$) with a flat radial density distribution ($q=0$, where $\rho[r]\propto r^{-q}$). The green tracks model clumpy geometries with varying numbers of dust clouds located along the line of sight ($N_0=1,3,5$ from upper right to lower left). The black star in the upper right corner is the starting point for all optically thin dust in the models ($S_{9.7}=1.26,S_{18}=0.67$). Silicate strengths from the IRS data are shown for the OHMS ({\it red}) and non-masing galaxies ({\it blue}). \label{fig-feature2}}
	\end{figure}

The infrared emission of ULIRGs is dominated by radiation from heated dust; thus, differences in the distribution of dust have significant influences on the mid-IR spectra. Since the mid-IR photons are responsible for maser pumping, this may also have an important effect on the presence of OHMs in ULIRGs. Using the radiative transfer code DUSTY, we modeled the dust environment of the galaxies in our sample for two geometries: a smooth, thick dust shell and a clumpy torus. 

The DUSTY code models the dust environment as a smooth, spherical distribution of centrally illuminated dust \citep{ive97}. Our models assumed the dust was composed of cool, oxygen-rich silicates \citep{oss92} and that the heating source follows a broken-power law luminosity function. The code fits for the thickness of the dust shell ($Y=R_{outer}/R_{inner}$), the power-law index $q$ of the radial density profile ($\rho[r]\propto~r^{-q}$), and the total optical depth $\tau_V$ at 0.55~\um. The inner dust radius, which is defined by the source luminosity and dust sublimation temperature, is the only free physical parameter in the code. DUSTY then generates a grid of artificial spectra at a variety of radii, from which parameters such as the dust temperature can be extracted. 

Motivated by the evidence that some fraction of our galaxies host AGN, we generated a second set of models for a clumpy distribution of dust, which may better represent the environment around active galaxies \citep{lan10}. We calculate the source function for individual clumps using DUSTY and used the code CLUMPY \citep{nen02,nen08,nen08a} to account for the new geometry. CLUMPY assumes a distribution of individual dusty clouds in a torus around the central illuminating source. The code fits for $Y$, $q$, and $\tau_V$ as well as the number of clouds along the line of sight ($N_0$), the angular dependence of cloud distribution away from the equatorial plane ($\sigma$), and the inclination angle of the galaxy ($i$). Neither $\sigma$ nor $i$ were well-constrained parameters in our models. 

We first used both the DUSTY and CLUMPY models to examine the overall dust distributions of the galaxies in our sample. The ``feature-feature'' diagram, developed by \citet{sir08}, plots the depths of the 9.7 and 18~\um~silicate features against each other. These are compared to tracks from the radiative transfer models; we used a small set of dust geometries and plotted the expected silicate ratios for a large range of optical depths. Following \citet{sir08}, we generated tracks for three smooth geometries that vary in shell thickness ($Y=100,200,400$ for $q=0.0$) and three clumpy geometries that vary in the number of clouds ($N_0=1,3,5$ for $q=0$, $Y=30$); the optical depth is then allowed to vary for each model from 0 to 80.

Figure \ref{fig-feature2} shows the tracks for the different dust geometries, as well as the measured silicate ratios for the OHMs and non-masing galaxies from the IRS spectra. The non-masing galaxies occupy a much smaller locus of possible dust geometries than the OHMs, showing no deep absorption ($S<-1.2$) in either silicate feature. As a result, most non-masing galaxies are best fit by one of the clumpy dust geometries. The OHMs occupy a much larger region; while a few galaxies fall close to the clumpy tracks, the majority of OHMs have deep 9.7~\um~absorption that only be achieved with a smooth, embedding medium. \citet{lev07} show that such deep absorption requires a large temperature gradient across the absorbing medium, which can only be achieved if the dust screen is both geometrically and optically thick. While the silicate ratios are not sensitive enough to strongly constrain either $Y$ or $N_0$, it does demonstrate a clear difference in the dust environments of the two ULIRG populations. 

Puzzlingly, most of the OHMs fall below the tracks predicted for the smooth dust geometries at $S_{9.7}<-1.5$. If this is a systematic effect, then this implies that either the 18~\um~feature is being overestimated or the 9.7~\um~feature underestimated with respect to the models. We consider the former more physically probable; if the OHMs were shifted to the left in Figure~\ref{fig-feature2} to lie on the smooth tracks, this would imply absorption depths of up to $S_{9.7}\simeq6-7$, much deeper than any seen in a ULIRG to date. As a check, we directly compared our measured silicate depths to those published in \citet{sir08} for ten galaxies that appear in both samples; however, no systematic difference in absorption strengths was found. Assuming the methods are consistent, this may indicate that a different geometry must be implemented in the code for the most heavily embedded ULIRGs. 

	\begin{figure*}
	\includegraphics[scale=1.0]{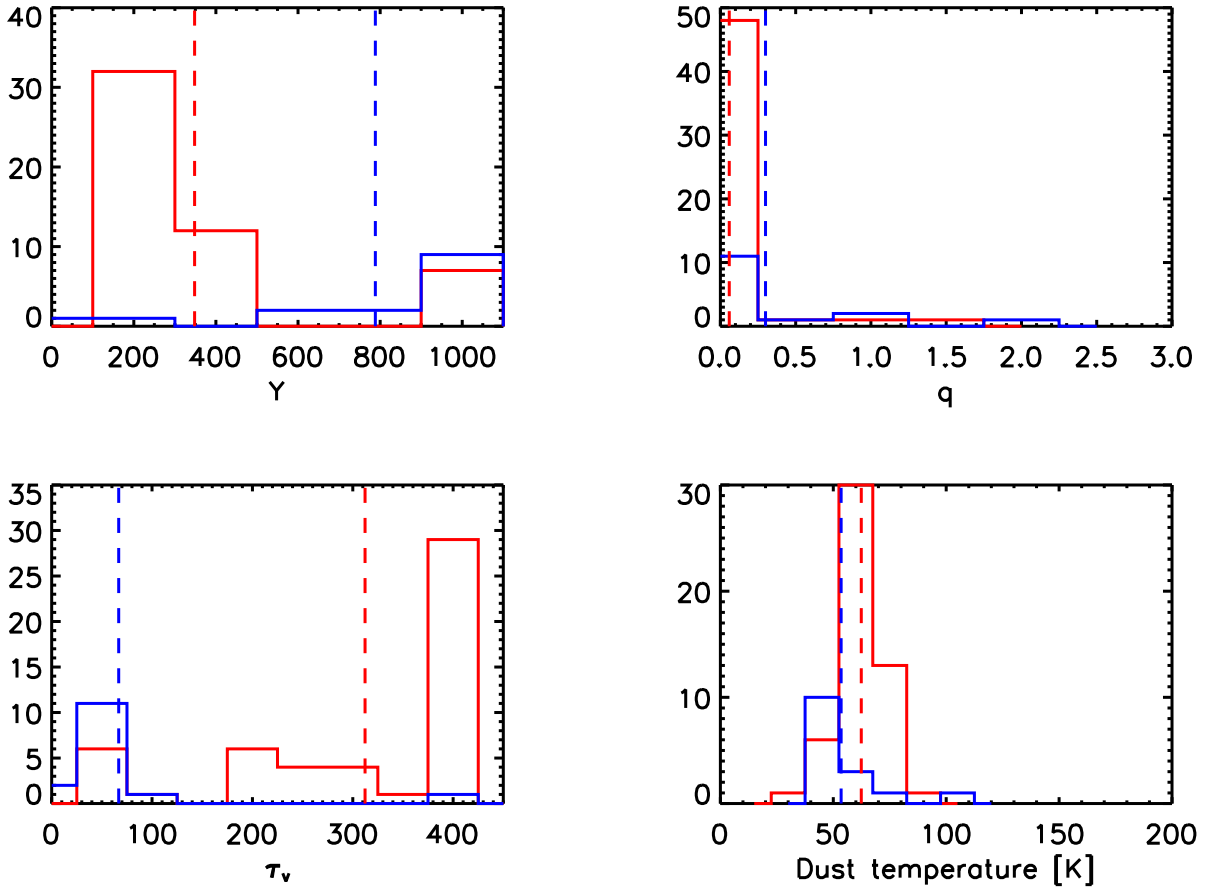}
	\caption{Distribution of the dust geometry parameters from DUSTY for the best fits to the IRS spectra. We modeled: ({\it top left}) the thickness of the dust shell ($Y=R_{outer}/R_{inner}$), the radial power-law index of the dust density $q$ ({\it top right}), the total optical depth $\tau_V$ ({\it bottom left}), and the dust temperature at the outer edge ({\it bottom right}). OHMs are in red and non-masing galaxies in blue, with the mean values for each samples indicated with dashed lines. \label{fig-dusty_pahfit_hist}}
	\end{figure*}
	
Figure~\ref{fig-feature2} suggests that OHMs are best modeled by a smooth geometry; based on these results, we attempted to further constrain the dust geometry for our galaxies by fitting each IRS spectra with the smooth-shell model. We used DUSTY to generate a grid of 13860 artificial spectra; the parameter ranges are in Table~\ref{tbl-grid}. We chose these values to span the expected physical range for ULIRGs: these include shell thicknesses out to $Y=1000$ (400~pc for typical values of dust sublimation temperature and the heating source luminosity), power-law indices from $0-3$, and $\tau_V$ extending up to 500.  

\begin{deluxetable}{ll}
\tabletypesize{\scriptsize}
\tablecaption{Grid parameters for DUSTY model fits\label{tbl-grid}}
\tablewidth{0pt}
\tablehead{
\colhead{} &
\colhead{DUSTY grid}
}
\startdata
$Y$      & 2, 5, 10, 15, 20, 30, 50, 75, 100, 150, 200, 300, \\ 
         & 400, 500, 750, 1000 \\ 
$q$      & 0, 0.5, 1.0, 1.5, 2.0, 2.5, 3.0    \\ 
$\tau_V$ & $0.1-500$                          \\ 
\enddata
\tablecomments{$\tau_V$ is binned on a logarithmic scale with 180 steps between 0.1 and 500.}
\end{deluxetable}

After generating the grid of artificial spectra, we needed to identify the best fit for each IRS spectrum. Since DUSTY only models continuum and dust features, we removed the PAH, atomic, and molecular lines from the IRS data to improve the quality of the fit. For this we employed PAHFIT, a set of IDL routines that performs spectral decomposition of low-resolution IRS data \citep{smi07}. While removing narrow line emission was typically clean, subtraction of the PAH emission often increased the area of the 9.7~\um~silicate feature, since the wings of the 8.6 and 11.3 PAH profiles fill in the dust absorption. Once the IRS spectra were reduced to continuum + dust features, we re-binned the data to the resolution of the DUSTY grid and found the best fit following the error minimization technique of \citet{nik09}. Results of the best fit $Y$, $q$, $\tau_V$, and $T_{dust}$ at the outer edge for each galaxy are given in Table~\ref{tbl-derived}. 

	Figure~\ref{fig-dusty_pahfit_hist} shows the distributions of the model parameters for the best DUSTY fits to the OHMs and non-masing galaxies. The best fits have a uniformly flat density profile for almost all galaxies in our sample; only 3/51 OHMs and 4/15 non-masing galaxies had best fits with $q>0$. The two samples also have similar dust temperatures, with OHMs slightly warmer on average than the non-masing galaxies ($\langle T_{dust}\rangle=62$ vs. 53~K), confirming the results of \citet{dar02}. These values are consistent with the greybody dust temperature measured with IR photometry (Equation~\ref{eqn-dusttemplate}), where $T_{dust}\sim45-75$~K for OHMs and $\sim40-120$~K for non-masing galaxies. 

	In contrast, the best fits for both the dust shell thickness ($Y$) and optical depth ($\tau_V$) are markedly different for OHMs and non-masing galaxies. The mean $Y$ for non-masing galaxies ($770\pm320$) is nearly twice as thick as the mean value for OHMs ($350\pm260$), although within the large scatter on both parameters. Rather than a physical difference in the shell thicknesses, however, we reiterate that this is a likely consequence of the non-masing galaxies being better fit by clumpy models (Figure~\ref{fig-feature2}) and thus a fundamentally different geometry. The dust optical depths for the OHMs have a broad distribution of $\tau_V$ between 0 and 450, with more than 50\% having $\tau_V>350$ and a mean of 300. With the exception of a single galaxy with $\tau_V=380$ (IRAS~06538+4628), all non-masing galaxies have $\tau_V<100$ and a median of less than 50. 

	Overall, fits to the IRS from radiative transfer models show a marked difference between the dust geometries of masing and non-masing galaxies. Non-masing galaxies are better fit by models with lower optical depth and slightly cooler dust temperatures - based on Figure~\ref{fig-feature2}, this may be due to dust in the form of a clumpy, obscuring torus (and possibly a partially visible AGN). OHMs are almost all well fit by a smooth screen of dust with a thinner shell, but with much higher optical depths. We emphasize that this ``smoothness'' is in the context of the entire nuclear region (and possibly beyond) of the merging galaxies. Smaller overdensities within that smooth framework are likely sites of star formation and are necessary to provide the cloud-cloud overlap that produces an OHM. 


\section{Statistical comparisons}\label{sec-stats}

	The primary goal in observing non-masing ULIRGs was to directly compare the samples and identify differences that could be triggers of the OHM. Here, we present statistical tests comparing the mid-IR and radio properties of both samples. 

\subsection{Rank correlations of IR and radio properties}\label{ssec-corr}

\begin{deluxetable*}{lrrrrrrrrrrrrrrr}
\tabletypesize{\scriptsize}
\tablecaption{Spearman rank correlation $z$-scores for OHMs\label{tbl-corr}}
\tablewidth{0pt}
\tablehead{
\colhead{}                   & 
\colhead{}                   & 
\colhead{}                   & 
\colhead{}                   & 
\colhead{}                   & 
\colhead{}                   & 
\colhead{}                   & 
\colhead{}                   & 
\multicolumn{3}{c}{\underline{DUSTY model fits}} &
\multicolumn{3}{c}{\underline{OHM radio properties}}
\\
\colhead{}                   & 
\colhead{$\alpha_{15-6}$}    &
\colhead{$\alpha_{30-20}$}   &
\colhead{$EW_{6.2}$}         &
\colhead{log $L_{6.2}$}      &
\colhead{$S_{9.7}$}          &
\colhead{$S_{18}$}           &
\colhead{$T_{dust}^{gb}$}    &
\colhead{$Y$}                &
\colhead{$\tau_{V}$}         &
\colhead{$T_{dust}$}         &
\colhead{log $P_{1420}$}     &
\colhead{log $P_{1667}$}     &
\colhead{log $L_{OH}$}       
}
\startdata

log $L_{FIR}$      &    0.1    &   $-$0.6   &    $-$1.4   &      1.8   &     $-$0.4   &     $-$0.2    &    $-$0.5   &       0.0   &       0.9    &    $-$0.5   & {\bf 4.4}    &        1.5  &      2.0   \\
$\alpha_{15-6}$    &    --     &   $-$0.4   &       1.4   &   $-$1.4   &        0.4   &     $-$0.4    &       1.0   &    $-$2.9   &       2.5    &       2.5   &       0.1    &     $-$0.8  &   $-$1.9   \\
$\alpha_{30-20}$   &  \ldots   &     --     &      2.3    &   $-$1.0   &    $-$2.3    &    $-$2.8     & {\bf $-$5.1}&      2.2    &      3.3     &   $-$2.7    &   $-$2.9     &       1.0   &  $-$0.0    \\
$EW_{6.2}$         &  \ldots   &   \ldots   &      --     &     2.5    &    $-$2.5    &    $-$1.9     &   $-$2.0    &   $-$1.5    &      3.7     &      0.9    &   $-$0.5     &       0.1   &  $-$0.1    \\
log $L_{6.2}$      &  \ldots   &   \ldots   &    \ldots   &     --     &    $-$3.6    &    $-$2.6     &      2.1    &   $-$1.0    &      0.5     &      0.8    &      2.5     &       1.7   &     2.4    \\
$S_{9.7}$          &  \ldots   &   \ldots   &    \ldots   &   \ldots   &       --     &  {\bf 5.3}    &      0.9    &      2.0    &   $-$3.5     &   $-$1.5    &      0.1     &    $-$2.9   &  $-$2.0    \\
$S_{18}$           &  \ldots   &   \ldots   &    \ldots   &   \ldots   &     \ldots   &       --      &      0.7    &      0.9    &   $-$2.9     &   $-$0.4    &   $-$0.0     &    $-$2.3   &  $-$2.1    \\
$T_{dust}^{gb}$    &  \ldots   &   \ldots   &    \ldots   &   \ldots   &     \ldots   &     \ldots    &      --     &   $-$1.7    &   $-$2.2     &      1.9    &      1.6     &    $-$0.3   &  $-$0.1    \\
$Y$                &  \ldots   &   \ldots   &    \ldots   &   \ldots   &     \ldots   &     \ldots    &    \ldots   &      --     &   $-$2.7     &{\bf $-$6.9} &   $-$0.5     &    $-$0.1   &     0.1    \\
$\tau_{V}$         &  \ldots   &   \ldots   &    \ldots   &   \ldots   &     \ldots   &     \ldots    &    \ldots   &    \ldots   &      --      &      1.5    &   $-$0.7     &       2.1   &     0.7    \\
$T_{dust}$         &  \ldots   &   \ldots   &    \ldots   &   \ldots   &     \ldots   &     \ldots    &    \ldots   &    \ldots   &    \ldots    &      --     &      0.4     &    $-$0.6   &  $-$0.6    \\
log $P_{1420}$     &  \ldots   &   \ldots   &    \ldots   &   \ldots   &     \ldots   &     \ldots    &    \ldots   &    \ldots   &    \ldots    &    \ldots   &      --      &       1.0   &     2.7    \\
log $P_{1667}$     &  \ldots   &   \ldots   &    \ldots   &   \ldots   &     \ldots   &     \ldots    &    \ldots   &    \ldots   &    \ldots    &    \ldots   &    \ldots    &       --    & {\bf 5.3}  \\
\enddata
\tablecomments{The Spearman's rank correlation tests statistical dependence of two variables on being monotonic functions, without assuming linearity. The $z$-scores in this table represent the number of standard deviations by which the correlation differs from the null hypothesis of no statistical dependence. Correlations higher than 4$\sigma$ are in boldface.}
\end{deluxetable*}

A first-order method of comparing the samples is correlation between physical parameters. We computed the Spearman's rank correlation coefficient ($\rho$) for a range of properties from Paper~I both in the mid-IR and radio regimes. For the mid-IR, we tested relationships between PAH 6.2 and 11.3~\um~EW and luminosities, continuum spectral indices, silicate absorption depths, greybody $T_{dust}^{gb}$, and constraints on the geometry from the DUSTY best-fit models (Table~\ref{tbl-derived}). The radio properties we explored included the continuum power at 1420~MHz ($P_{1420}$), integrated $L_{OH}$ (Paper~I) and the peak OHM flux density at 1667~MHz \citep[$P_{1667}$;][]{dar00,dar01,dar02}. Results for the Spearman's $\rho$ tests are shown in Table~\ref{tbl-corr}. We omit several parameters from this table that were measured, but showed no significant correlations; these included the mid-IR fine-structure line ratios (\neIII/\neII, \oIV/\neII, and \neV/\neII), \htwo~temperature and gas mass, depth of the 6~\um~water ice feature, and the OH hyperfine ratio $R_H=F_{1667}/F_{1665}$. 

Several of the correlations in Table~\ref{tbl-corr} with high significance reflect well-known physical relationships; for example, the correlation between $P_{1420}$ and $L_{FIR}$. The relation of the spectral index $\alpha_{30-20}$~to $T_{dust}^{gb}$ is expected since $\alpha_{30-20}$ samples the Wien side of a blackbody peaking near 60~\um. The silicate depths at 9.7 and 18~\um~are also correlated, as expected from the results of the DUSTY models. 

The dust temperature and shell thickness $Y$ from the DUSTY best-fit models showed a strong anti-correlation in OHMs. This picture fits with a smooth dust screen enveloping a central source of illumination - thicker shells absorb more energy near the inner boundary, resulting in a cooler $T_{dust}$ near the outer boundary. $Y$-$T_{dust}$ also was the only correlation coefficient that showed significant differences between the masing and non-masing samples ($\rho_{OHM}=-0.97$,~$\rho_{non}=-0.64$). We attribute this to the evidence that non-masing galaxies are poorly fit by DUSTY and likely favor a clumpy geometry (Figure~\ref{fig-feature2}). 

For the OH maser emission itself, the only strong correlation observed was between the peak OHM power and the integrated OHM luminosity. No Spearman correlations between an OHM parameter ($L_{OH}$, $P_{1667}$) and the IRS data were found with $>4\sigma$ significance, despite the fact that several IR features related to dust geometry revealed a clear separation of loci between OHMs and non-masing galaxies (see Figures~\ref{fig-spoon_forkII} and \ref{fig-sil_a3020}). While these parameters have a clear effect on the {\em existence} of the OHM, the lack of correlation suggests either that specific line properties are not well tracked by these parameters, or that no single trigger among these is responsible for megamaser production. Alternatively, the OHM might be the results of stochastic amplification of small-scale conditions, with masing simply becoming more common when conditions are favorable. 

\subsection{Statistical differences between the samples}\label{ssec-statdiff}

\begin{deluxetable}{lccrrrr}
\tabletypesize{\scriptsize}
\tablecaption{Kolmogorov-Smirnov tests for OHMs and non-masing galaxies\label{tbl-ks}}
\tablewidth{0pt}
\tablehead{
\colhead{} &
\colhead{$D_{KS}$} &
\colhead{$N_{\sigma}$} &
\colhead{$\mu_{OHM}$} &
\colhead{$\sigma_{OHM}$} &
\colhead{$\mu_{non}$} &
\colhead{$\sigma_{non}$} 
}
\startdata										
$EW_{6.2}$ [\um]            & 0.440 &     2.4  & $9.4$   &  47.5  & $1.13$  & 1.32  \\   
log $L_{6.2}$ [$L_\sun$]    & 0.193 &     0.3  & $9.70 $ &  0.38  & $9.63 $ & 0.48  \\   
log $L_{FIR}$ [$L_\sun$]    & 0.400 &     2.1  & $12.18$ &  0.27  & $11.85$ & 0.52  \\   
$f_{60}/f_{100}$            & 0.318 &     1.2  & $-0.08$ &  0.12  & $-0.15$ & 0.12  \\   
$\alpha_{15-6}$             & 0.326 &     1.4  & $2.08 $ &  0.61  & $1.80 $ & 0.58  \\   
$\alpha_{30-20}$            & 0.816 & {\bf5.3} & $4.89 $ &  1.08  & $2.50 $ & 0.89  \\   
$S_{9.7}$                   & 0.800 & {\bf5.1} & $-1.83$ &  0.76  & $-0.62$ & 0.37  \\   
$S_{18}$                    & 0.686 & {\bf4.4} & $-0.56$ &  0.31  & $-0.23$ & 0.13  \\
$T_{gb}$ [K]                & 0.369 &     1.9  & $65   $ &  11    & $79   $ & 49    \\   
$T_{30-20}$ [K]             & 0.882 & {\bf5.8} & $75   $ &  14    & $113  $ & 29    \\   
$T_{DUSTY}$ [K]             & 0.663 & {\bf4.2} & $62   $ &  10    & $53   $ & 19    \\   
$\tau_{V}$                  & 0.816 & {\bf5.3} & $310  $ &  130   & $67   $ & 88    \\   
$Y$                         & 0.729 & {\bf4.7} & $350  $ &  250   & $790  $ & 320   \\   
$q$                         & 0.518 &     3.0  & $0.06 $ &  0.26  & $0.30 $ & 0.59  \\   
\enddata
\tablecomments{The parameters $\tau_V$, $T_{DUSTY}$, $Y$, and $q$ are the best fit of IRS spectra to DUSTY models. Results greater than 4$\sigma$ significance are in boldface.}
\end{deluxetable}

A second method of analyzing statistical differences between OHMs and non-masing galaxies is the two-sided Kolmogorov-Smirnov (K-S) test, which tests the null hypothesis that the two samples come from the same parent distribution. Selected results from the K-S tests are given in Table~\ref{tbl-ks}, where $D_{KS}$ is the maximum separation between the scaled cumulative distribution functions and $N_\sigma$ is the number of standard deviations by which $D_{KS}$ differs from the null hypothesis (ie, the significance of the result). We also give the mean values and $1\sigma$ standard deviations of the properties for both samples. As with Table~\ref{tbl-corr}, this omits numerous mid-IR and radio properties on which we performed K-S tests, but which showed no significant difference. 

The majority of the data show K-S results consistent with origins from the same distribution; the exceptions all relate to dust properties of the galaxies. Two of these are quantities directly measured from IRS data: the 30-20~\um~spectral index and the 9.7~\um~silicate depth. The remaining significant parameters describe the dust environment modeled by DUSTY: $Y$, $\tau_V$, and $T_{dust}$ all support a fundamentally different distribution of the silicate dust for the two samples at the 4$\sigma$ level. 

We extended this analysis by refining the greybody $T_{dust}$ measured with Equation~\ref{eqn-dusttemplate}, which shows only a mild significance in the original K-S test ($2\sigma$). The continuum slopes are largely determined by the amount of dust in various temperature regimes; however, $\alpha_{30-20}$ shows a strong difference while $\alpha_{15-6}$ (which samples hotter dust) does not. This may indicate that only dust {\em in certain temperature regimes} (ie, the $\sim50-100$~K region sampled by 20--30~\um~continuum) is important in triggering OHM emission. We tested this by restricting the fit of our dust temperature only to data from 20--30~\um, where the Wien approximation applies for typical ULIRG dust temperatures. In this case, the K-S test yields a much higher and statistically significant ($6\sigma$) difference for the modified dust temperature ($T_{30-20}$) between the OHMs and non-masing galaxies. 

K-S tests were also used to quantify the differences seen between the samples on the fits with the DUSTY code. In particular, the optical depth $\tau_V$ showed a $5\sigma$ difference between the two samples, with the typical OHM having $\tau_V$ a factor of several above a non-masing galaxy (and consistent with $D_{KS}$ for the 9.7~\um~feature). The other DUSTY parameters show moderate significance (3--5$\sigma$); however, the results for different distributions of $Y$ likely come from a clumpy geometry, rather than a true increase in the shell thickness (\S\ref{ssec-dustgeometry}). Given that the optical depth and dust temperature are {\em not} independent parameters in DUSTY, all results from the K-S tests strongly indicate that the temperature/optical depth of the dust (which depends on its geometry) is a key factor in triggering an OHM. 

We extended the K-S tests by performing a series of survival analyses on the same data. Survival analysis is particularly suited for flux-limited samples because it properly treats upper limits for features not detected in all galaxies \citep{fei85,iso86}. We used the ASURV package in IRAF \citep{lav92}, which includes the Gehan's generalized Wilcoxon, logrank, Peto \& Peto, and Peto \& Prentice tests. Running survival analysis on all measured mid-IR features (including atomic and molecular line emission, hydrocarbon and gas-phase absorption, PAH, dust and continuum features) gave similar results to the K-S tests; no parameter showed significant differences between the two samples with the exceptions of $\alpha_{30-20}$ and $S_{9.7}$. All tests yielded statistically significant differences for these features, with an mean significance of $6\sigma$ for $S_{9.7}$ and $5\sigma$ for $\alpha_{30-20}$. 

Importantly, the results of our survival analysis also discount the possibility that other mid-IR features are directly related to the presence of an OHM. In particular, we detected absorption from hydrocarbons (HACs), gas-phase molecules (\acet, \hcn, and \cotwo) and crystalline silicates almost exclusively in the OHM sample (Paper~I). Using the upper limits on the absorption features in our survival analyses, however, we cannot confirm that the lack of detections in the non-masing sample is significant above the $3\sigma$ level for any of these parameters. This is largely due to lack of sensitivity in the non-masing galaxies, since the limits on non-detections are of similar magnitudes to the detected absorption in many galaxies. 


\section{Comparing observations and theory}\label{sec-theory}

	Importantly, the IRS data can explore the physics of OHMs by testing the predictions of maser pumping models. The most recent and complete pumping calculations come from \citet[][hereafter LE08]{loc08}. The model assumes a slab geometry and uses the escape probability method to solve for the level populations of the OH molecule. Given assumptions on the physical conditions in the masing regions, the overall strength of the OHM (if any) can be predicted. The clumpy OHM model of \citet{par05} shows that the maser optical depth depends most strongly on the dust temperature and optical depth. Since both these parameters can be estimated from IRS data, our sample offers the first opportunity for testing such a model on a large number of galaxies. 

	The strength of the OHM in the LE08 model is parameterized as the optical depth in the OH line ($\tau_{1667}$, which becomes more negative for higher maser gain). To compare this to observations, we use the line-to-continuum ratio to estimate the apparent observed OH optical depth (Table~\ref{tbl-derived}):

\begin{equation}
\label{eqn-tauohapp}
\tau_{1667}^{app} = -\textrm{ln}\left(\frac{S_{1420} + S_{1667}}{S_{1420}}\right).  
\end{equation}

\noindent Here $S_{1667}$ is the peak flux density of the OHM at 1667~MHz \citep[taken from][]{dar00,dar01,dar02} and $S_{1420}$ is the flux density of the radio continuum at 1420~MHz from the NRAO VLA Sky Survey \citep{con98}. 

	The largest uncertainty in Equation~\ref{eqn-tauohapp} is that it assumes an OH filling factor of 1; however, VLBI maps of OHM galaxies show the OH emission to have both diffuse and compact components. In addition, the 1420~MHz radio continuum comes from a much larger physical area than the OH emission. Therefore, $\tau_{1667}^{app}$ will be a weaker limit to the true 1667~MHz optical depth. VLBI observations have mapped the OHM emission for a handful of nearby galaxies \citep{yat00,pih01,pih05,klo03,lon03,rov03,klo04,ric05,mom06}, showing that the difference in apparent optical depth between the entire galaxy and the brightest individual maser spots varies by as much as $\Delta\tau\simeq1-4$. Furthermore, the gain for individual maser spots with cloud-cloud overlap can be as high as several hundred \citep[eg, III~Zw~35;][]{dia99,par05}, compared to the diffuse background. Since high-resolution OH maps do not exist for the vast majority of the IRS galaxies, however, we use $\tau_{1667}^{app}$ while remaining mindful of the above caveats. 

We present tests of the LE08 model parameters using two different techniques: one method estimating $T_{dust}$ and $\tau_V$ directly from the IRS spectra, and a second using parameters extracted from the models fit to the IRS data using the DUSTY code. 

\subsection{Testing the LE08 pumping model with $T_{dust}$ and $\tau_V$ from IRS data}\label{ssec-le08v1}

The pumping flux of the OHM in the LE08 model depends most strongly on the pumping flux, which is controlled by two factors: the self-absorption of the dust (depending on $\tau_V$) and the Planck function (depending on $T_{dust}$). The 9.7~\um~silicate feature in the IRS data can be used to estimate the total $\tau_V$ using the Galactic calibration of \citep{roc84}:

\begin{equation}
\label{eqn-sil_tauv}
\tau_V = (17.0\pm1.4) \times \tau_{9.7}.
\end{equation}

\begin{figure}
\includegraphics[scale=0.5]{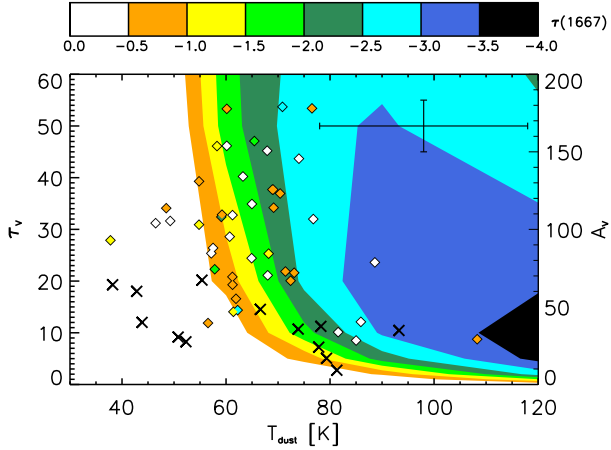}
\caption{Apparent optical depth of the OHM emission (\textit{filled diamonds}) as a function of $\tau_V$ (Equation~\ref{eqn-sil_tauv}) and dust temperature (Table~\ref{tbl-derived}), based on greybody photometry and the 9.7~\um~silicate depth. Crosses show temperatures and depths of the non-masing sample. The error bars in the top left are the average systematic uncertainties ($\sigma_T=20$~K, $\sigma_\tau=5$). Contours are a zoomed-in region of the OHM model of \citet{loc08}. \label{fig-lockettplot}}
\end{figure}

	Secondly, the dust temperature in the masing region is estimated from the greybody fit to the IRS and IRAS data (\S\ref{ssec-dusttemp}; Table~\ref{tbl-derived}). Once both parameters for a galaxy are estimated, we plot the data on the contours of LE08-predicted $\tau_{1667}$ emission (Figure~\ref{fig-lockettplot}). The color of the OHM symbols shows their $\tau_{1667}^{app}$ on the same scale as the LE08 contours. 

	According to the standard LE08 model, the most luminous OHMs are expected to have $T_{dust}\sim90-150$~K and $\tau_V$ of a few tens. None of the OHMs with $\tau_{1667}^{app}<-3.0$ were located near this region; in addition, the parameters for virtually all observed OHMs lie well away from the highest predicted $\tau_{1667}$ in Figure~\ref{fig-lockettplot}, with $T_{dust} = 40-100$~K and $\tau_V = 10-50$. Roughly 15\% of the confirmed OHMs have predicted $\tau_{1667}>-0.5$, which would predict almost no masing activity and would fall well below the limit for inclusion in our sample. 

	To assess the overall fit of the LE08 model, the errors on our estimates of $\tau_V$ and $T_{dust}$ must be quantified.  \citet{yun02} found that accurate temperature fits using Eqn.~\ref{eqn-dusttemplate} required much higher photometric sampling than we possess for this sample, including radio and sub-mm data. Other SED models \citep{fra99,dun01} typically fit two- and three-component models with differences in dust temperatures ranging up to 100~K. As a result, we estimate the average $\sigma_{T_{dust}}$ for each galaxy to be $\sim20$~K. This is quite high, but is mitigated somewhat by the number of galaxies in our sample. The mean error on the extinction is estimated as $\sigma_{\tau_V}=5$, based on the silicate measurement technique, the possibility of saturation, and calibration in Eqn.~\ref{eqn-sil_tauv}. Average error bars are shown in the upper right corner of Fig.~\ref{fig-lockettplot}.  

\begin{figure}
\includegraphics[scale=0.5]{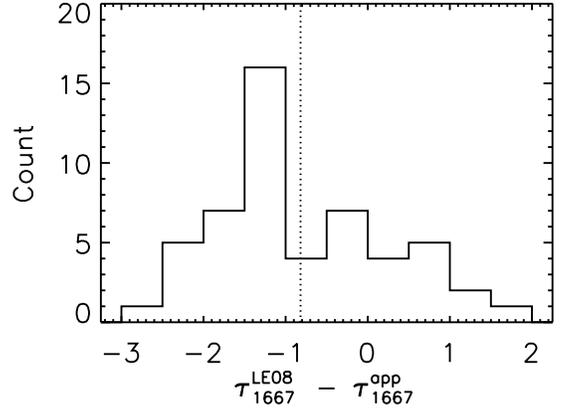}
\caption{Distribution of the difference in Figure~\ref{fig-lockettplot} between the predicted $\tau_{1667}$ from the \citet{loc08} model and the apparent $\tau_{1667}^{app}$ measured from radio data. The dotted line shows the mean of the distribution at $\Delta\tau_{1667}=-0.8$, showing that the LE08 model tends to overpredict the strength of the maser. \label{fig-locketthist}}
\end{figure}

	The large uncertainties result in considerable scatter in the predicted OHM strength, with $\Delta\tau_{1667}$ as high as 1--2 depending on the local gradient of the LE08 model. Many of the OHMs lie near contours where $\tau_{1667}$ is a sensitive function of $T_{dust}$; a shift of $\sim5$~K could result in a change of up to $\Delta\tau_{1667} = 0.5$, while at the same time being relatively insensitive to $\tau_V$. We show the distribution of the difference between $\tau_{1667}^{app}$ and the predicted $\tau_{1667}$ from the LE08 model in Figure~\ref{fig-locketthist}. The measured OH strengths are on average weaker than those predicted by the model ($\langle\Delta\tau_{1667}\rangle=-0.8$); this is consistent, however, with the lower bound on $\tau_{1667}^{app}$ from the OH filling factor. The $\chi^2_{reduced}$ for the model using this data is 29.1, which rejects a correlation hypothesis at the 5$\sigma$ level. 

While the agreement for individual OHMs is not strong, our data is consistent with other predictions of the LE08 model. Based on pumping calculations, they show that ULIRGs must have a dust temperature greater than $45$~K in order to achieve population inversion; cooler temperatures move the peak of the blackbody too far from the main pumping lines to support the necessary pumping flux. 90\% of the OHMs have $T_{dust}>45$~K, with the coolest OHM at 37~K; uncertainties of $\sim20$~K mean that the dust temperatures for all OHMs are fully consistent with this predicted lower limit. 

	Figure~\ref{fig-lockettplot} also displays the dust parameters for the non-masing galaxies. Half of the non-masing galaxies have predicted OH luminosities consistent with little masing to none at all (such that $|\tau_{1667}|$ is small), and the $\tau_{1667}$ predicted by the LE08 model would lie below our detection threshold of $L_{OH}\le10^{2.3}$~L$_\sun$ for almost the entire sample. Based on the LE08 model and IRS data, only a single non-masing galaxy (IRAS~23498+2423, in the far lower right of Fig.~\ref{fig-lockettplot}) would have been expected to show strong megamaser emission. Interestingly, IRAS~23498+2423 was the object with the {\em highest} upper limit on maser emission in the non-masing sample, with $L_{OH}<2.25$. We re-observed this galaxy at the Arecibo Observatory\footnote{The Arecibo Observatory is part of the National Astronomy and Ionosphere Center, which is operated by Cornell University under a cooperative agreement with the National Science Foundation.} in October 2009 to test the LE08 prediction; no detection of OH was made, confirming an upper limit of $L_{OH}\le10^{2.27}$~L$_\sun$. 


\subsection{Testing the LE08 pumping model with fits to IRS data from DUSTY}\label{ssec-le08v2}

	Our second approach for testing the LE08 model calculates $T_{dust}$ and $\tau_V$ from the best fits to the IRS data with DUSTY (\S\ref{ssec-dustgeometry}; Table~\ref{tbl-derived}). As discussed in \S\ref{ssec-statdiff}, these fits give dust temperatures similar to those from the greybody fit, but with optical depths much higher than those calculated using only the 9.7~\um~feature, which can be saturated in ULIRGs. The $T_{dust}$ used here is the value at the shell's outer edge; since the radial temperature profile of the dust is very steep close to the center and shallow at the edges (changing by only a few tens of K over the outer half of the shell), this temperature represents the bulk of the dust mass and is likely a reasonable approximation for conditions in the masing regions. 

\begin{figure}
\includegraphics[scale=0.5]{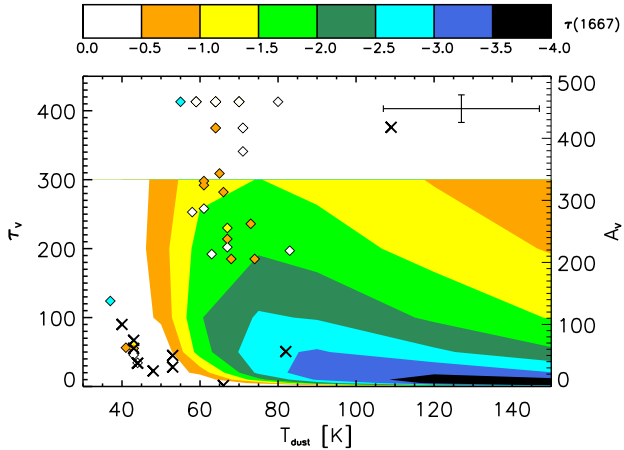}
\caption{Apparent optical depth of the OHM emission (\textit{filled diamonds}) as a function of $\tau_V$ and dust temperature based on fits from DUSTY. Black crosses are the non-masing galaxies; the error bar in the upper right corner shows the average uncertainty ($\sigma_\tau\sim20,\sigma_T\sim20$~K). Contours show the predicted maser strength for the LE08 model, which are not complete at $\tau_V>300$. Due to the coarse gridding of DUSTY models, several galaxies have overlapping points on this plot (eg, five galaxies have best fits of $T_{dust}=70$~K and $\tau_V=410$). \label{fig-lockettplot_dusty}}
\end{figure}

	Figure~\ref{fig-lockettplot_dusty} shows the LE08 predictions with data from the DUSTY best fits to the OHMs and non-masing galaxies. We note that LE08 contours are not complete at $\tau_V>300$, and that the apparent horizontal feature at $\tau_V=410$ is likely an artifact of gridding in the code. The distribution of the galaxies is very different from that in Figure~\ref{fig-lockettplot}; the OHMs occupy a much larger range in optical depth, increasing $\tau_V$ by an order of magnitude. Two distinct loci are visible; the lower left corner contains the majority of the non-masing galaxies and several OHM with $T_{dust}\simeq40-60$~K and $\tau_V<100$. The LE08 model predicts that these galaxies would show little to no maser emission. The second group is almost exclusively composed of OHMs, with warmer temperatures ($T_{dust}\simeq60-80$~K) and $\tau_V$ of several hundred. LE08 predicts a range of $\tau_{1667}$ for these galaxies from $-2.0$ to 0. Two non-masing galaxies with warm temperatures do not seem associated with either group. 

	The overlapping region of OHMs and non-masing galaxies with cooler dust temperatures and $\tau_V<150$ is of particular interest. The dust parameters for these galaxies lie well away from the predicted $\tau_{1667}$ peak and are close to the minimum predicted inversion temperature of 45~K. The overlapping populations at this locus include at least one powerful gigamaser, and imply that there must be some triggering factor for an OHM beyond $T_{dust}$ and $\tau_V$. This is supported by the fact that {\em all} OHMs in this region also lie on the horizontal branch of the fork diagram (Fig.~\ref{fig-spoon_forkII}), classifying them as likely AGN hosts. The $T_{dust}$ from DUSTY for all of these OHMs is also significantly cooler than that measured with the greybody method; a buried AGN might thus be better fit with a multi-temperature model. The connection between an AGN and OHM suppression, however, is not clear; it could represent a different dust geometry not conducive to cloud-cloud overlap, or signal a more advanced stage in the galaxy merger, thus putting a limit on the effective lifetime (and thus observability) of the OHM. 

	Since the contours in Figure~\ref{fig-lockettplot_dusty} are incomplete, we cannot fully measure the goodness-of-fit in a method similar to Figure~\ref{fig-locketthist}. Qualitatively, the model makes good predictions for the dust parameters for almost all of the non-masing galaxies. The exceptions are IRAS~11119+3257 (log~$\tau_{1667}^{pred}\simeq-3.0$) and IRAS~06538+4628 (log~$\tau_{1667}^{pred}\simeq-1$). The former is the only object whose predicted emission lies well above its observational limits on $L_{OH}$. This galaxy is known to show an exceptional radio excess as measured by its $q$-parameter \citep{con91}, with its value of $q=1.23$ falling well below the mean value for OHMs found by \citet{dar02}. Such an excess commonly indicates that the galaxy hosts an AGN; this is supported by the low 6.2 PAH EW of IRAS~11119+3257 and its position on the fork diagram (Fig.~\ref{fig-spoon_forkII}).  IRAS~06538+4628 is the only non-masing galaxy in our sample for which dust temperatures might be too warm to support a strong population inversion. 
	
	There are several reasons why the $T_{dust}$ and $\tau_V$ fit from DUSTY might differ from methods used in \S\ref{ssec-le08v1}. $\tau_V$ for deeply embedded galaxies can suffer from saturation in the 9.7~\um~feature. Using data from the full dust profile (as we do in DUSTY) samples the broader wings of the feature it advances up the curve of growth. DUSTY also samples the SED at a much higher resolution than the greybody fit, albeit in a more limited wavelength regime. Finally, the conversion from $S_{9.7}$ to $\tau_V$ is based on a Galactic calibration; it is not known how dust composition might be different in the Milky Way and ULIRGs, for example. On the other hand, the DUSTY models require the assumption of a specific geometry which may not be appropriate (see Fig.~\ref{fig-feature2}) for non-masing galaxies and the dustiest OHMs. While neither method is without drawbacks, we believe both to have at least some physical merit (and are encouraged by the fact that $T_{dust}$ is mostly consistent).
	
\subsection{Predictions and future observations}\label{ssec-predictions}
	
	Overall, comparing the IRS data to predictions from the LE08 model yielded mixed results. Using parameters from the DUSTY code, the LE08 model correctly predicted that most non-masing galaxies should have cool dust temperatures and low optical depths; however, several megamasers also have $T_{dust}$ that would be too cool for inversion under this model. The observed $\tau_{1667}^{app}$ for individual sources shows a great deal of scatter; however, this is dominated by observational uncertainties in the OH filling factor. Both estimates are consistent with the LE08 claim that a minimum dust temperature of 45~K is required for maser action; within uncertainties, all OHMs have $T_{dust}$ above this value. Based on results from the feature-feature diagram, we suggest that future pumping models include both clumpy and smooth shell dust geometries; treatment of OH kinematics might also be necessary to model individual sources in more detail. 

Interestingly, the OHM luminosity (which typically depends strongly on the total linewidth) does not appear to be a strong function of the currently observable global host properties. An improved test of the $L_{OH}$-host galaxy relationship could use VLBI maps of OHM galaxies to constrain the true gain in individual clouds to determine the filling factor, and then compare these results to spatially resolved IR data in the same regions to measure $T_{dust}$ and $\tau_V$. If the parameters for OHM production can be fine-tuned based on size scales of $\sim100$~pc for nearby galaxies, this will greatly assist in comparisons of galaxy-wide SEDs for OHMs to non-masing ULIRGs at much greater distances. 

The LE08 model depends on a number of other ISM properties, some of which can be further constrained by the IRS data. These include the ortho-para ratio of \htwo, which affects collision rates and thermalization of the gas. LE08 assumes a constant ortho-para ratio of 3; the IRS data show that this is only valid for 4/9 ULIRGs for which the ratio can be constrained (and can be as low as 0.5). The OH column densities measured using the 34.6~\um~transition lie in the range $N_{OH}=1-3\times10^{17}$~cm$^{-2}$; this is roughly a factor of two higher than the standard value assumed in the LE08 model. 

\begin{figure}
\includegraphics[scale=0.5]{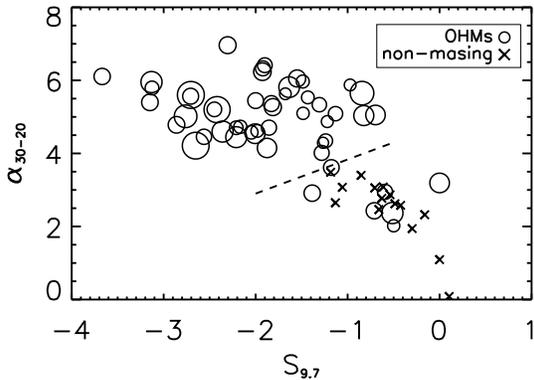}
\caption{Peak silicate depth at 9.7~\um~vs. the spectral slope between 20 and 30~\um~for OHMs ({\it circles}) and non-masing galaxies ({\it crosses}). For OHMs, the symbol size is proportional to log~$L_{OH}$. The dashed line shows the rough separation between the loci of OHMs and non-masing galaxies. Preselecting OHM hosts based on a $\alpha_{30-20}-S_{9.7}$ cut may be a powerful technique for future megamaser surveys.\label{fig-sil_a3020}}
\end{figure}

Results from our IRS data can also narrow potential searches for new OHMs, especially at higher redshifts. The most distant OHM known lies at $z = 0.265$ \citep{baa92}. Since OHMs are associated with merging galaxies, which are most plentiful between $z\sim1−-3$, we expect a higher spatial density of OHMs in the early universe \citep{dar02a}. While surveys for more distant OHMs are restricted both by sensitivity constraints and low-frequency RFI, a significant obstacle has been identification of a suitable target sample of host galaxies. Based on the IRS data, we suggest that future OHM surveys target galaxies with dust peaks near $\lambda_{rest}=53$~\um, steep $30-20$~\um~slopes, deep dust absorption, and that do not show evidence of hosting an AGN. Figure~\ref{fig-sil_a3020} shows how the combination of $\alpha_{30-20}$ and $S_{9.7}$ can clearly separate almost all OHMs from non-masing galaxies in the mid-infrared. This may be a valuable tool in future searches for OHMs; for galaxies in which low-resolution IR spectroscopy is available, pre-selecting OHM candidates based on these diagnostics should have a success rate far in excess of blindly selecting ULIRGs from the field. The growing number of sub-millimeter galaxy catalogs and multiwavelength deep fields offer excellent opportunities in the near future for such surveys.

Finally, new observatories are offering opportunities for completing the OHM picture. There are only a few OHMs in which the important 53~\um~transition has been measured \citep[eg,][]{he04}; this could be potentially obserbed in larger numbers of galaxies with SOFIA. The Herschel observatory can also supplement the IRS by measuring the 79 and 119~\um~OH transitions in large samples of ULIRGs \citep[eg,][]{fis10}. If models such as LE08 can be refined based on direct OH measurements, then the pumping efficiencies for the megamaser could be evaluated for a statistically significant sample. Photometric measurements from these instruments and from JWST will also generate SEDs with much broader spectral coverage, increasing our knowledge of the physics needed for radiative transfer models and modeling the OHM environment.


\section{Conclusions}\label{sec-conclusions}

We present results from the {\it Spitzer} Infrared Spectrograph comparing the mid-IR properties of OH megamaser hosts to galaxies with confirmed upper limits on the megamaser emission. No significant differences between the samples were found for the average excitation states, line velocities, or star formation rates. 10--25\% of the OHMs show clear evidence in the mid-IR for an AGN, significantly lower than previous optical and radio studies which placed the AGN fraction of OHMs between 30--70\%. In non-masing ULIRGs, between 50--60\% of the galaxies have mid-IR evidence for an AGN. 

Fits of radiative transfer models to the IRS spectra with the DUSTY code show that OHMs have warmer $T_{dust}$ and deeper silicate absorption associated with a smooth, thick dust shell surrounding the nucleus. This implies the presence of a large dust reservoir in OHMs with a smooth geometry and temperatures from $\sim50−-100$~K. Non-masing galaxies show weaker dust absorption, shallower mid-IR continuum, and cooler dust (by $\sim10$~K) than the typical OHM host. The relative strength of silicate features in non-masing galaxies suggests that they are best fit with a clumpy dust geometry. 

We used IRS data to evaluate predictions from the OH pumping model of \citet{loc08}, the first direct test of OHM production using observed properties of the host galaxies. The dust opacities for OHMs derived from the best-fit DUSTY models suggest that much higher opacities ($\tau_V\sim100-400$) are necessary for OHM production. All the IRS data are consistent with the LE08 claim that a minimum $T_{dust}=45$~K is required for maser action. Limits on the OH emission for most non-masing galaxies are predicted by the LE08 model, based on their comparatively cool dust temperatures ($T_{dust}<60$~K) and low dust opacity ($\tau_V<100$). Finally, the IRS data constrain several parameters necessary to develop future OHM pumping models, including the dust optical depth, temperature, and overall geometry.

For the first time, we present spectral diagnostics that can distinguish OHMs from non-masing galaxies based on their host galaxy properties, the clearest of which is the $S_{9.7}-\alpha_{30-20}$ relation (Figure~\ref{fig-sil_a3020}). These parameters can be relatively easily measured in low-resolution spectra, and may signify a powerful method for preselecting OHM candidates for follow-up radio surveys at higher redshift. 


\acknowledgements

This work is based on observations made with the \textit{Spitzer Space Telescope}, which is operated by the Jet Propulsion Laboratory (JPL) and the California Institute of Technology, under NASA contract 1407. Additional observations were also taken at the Arecibo Observatory (NAIC/Cornell). We made extensive use of the NASA/IPAC Extragalactic Database (NED) which is operated by JPL and Caltech under contract with NASA. Many thanks to are due to J. Stocke for jump-starting work with \textit{Spitzer}, D.~Dale for discussions about mid-IR line ratios, P.~Lockett and M.~Elitzur for sharing their radiative transfer models, and the Spitzer Science Center for hosting KW and JD while collaborating on data analysis. VC acknowledges partial support from the EU ToK grant 39965 and FP7-REGPOT 206469. 



\bibliography{kwrefs}

\end{document}